 \documentclass[preprint2]{aastex}

\shorttitle{NGC~376}
\shortauthors{Sabbi et al.}

\begin{document}


\title{Is the young star cluster NGC~376 dissolving in the field of the SMC? \altaffilmark{1}}


\author{E. Sabbi\altaffilmark{2}, A. Nota\altaffilmark{2,3}, M. Tosi\altaffilmark{4}, L.J. Smith\altaffilmark{2,3},  J. Gallagher\altaffilmark{5}, \& M. Cignoni\altaffilmark{4,6}}

\email{sabbi@stsci.edu}


\altaffiltext{1}{Based on observations with the NASA/ESA {\it Hubble Space Telescope}, obtained at the Space Telescope Science Institute, which is operated by AURA, Inc., under NASA contract NAS5--26555. These observations are associated with program 10248.} 
\altaffiltext{2}{Space Telescope Science Institute, 3700 San Martin Drive, Baltimore, MD 21218, USA}
\altaffiltext{3}{European Space Agency, Research and Scientific Support Department, Baltimore MD, USA}
\altaffiltext{4}{Istituto Nazionale di Astrofisica, Osservatorio Astronomico di Bologna, Via Ranzani 1, I-40127, Bologna, Italy}
\altaffiltext{5}{Department of Astronomy, University of Wisconsin, 475 North Charter St., Madison, WI 53706, USA}
\altaffiltext{6}{ Dipartimento di Astronomia, Universit\` a degli Studi di Bologna, via Ranzani 1, I-40127 Bologna, Italy}


\begin{abstract}

We use deep images acquired with the Advanced Camera for Surveys (ACS) on board of the {\it Hubble Space Telescope} (HST) in the filters F555W and F814W to characterize the properties of NGC~376, a young star cluster located in the wing of the Small Magellanic Cloud (SMC). Using isochrone fitting we derive for NGC~376 an age of $28\pm 7$ Myr, in good agreement with previous studies. The high spatial resolution ACS data allow us to determine the center of gravity of the cluster and to construct  extended surface brightness and radial density profiles. Neither of these profiles can be fitted with a theoretical model, suggesting that the cluster is not in virial equilibrium. Considering the young age of the cluster, we speculate that the distortion of the radial profiles may be the result of the rapid gas dispersal that follows the initial phase of star formation. The cluster shows clear evidence of dynamical mass segregation. From the properties of the radial profiles and the present day mass function (PDMF) we conclude that NGC~376 appears to have already lost nearly 90\% of its initial stellar mass, probably as a consequence of the sudden gas dispersal that follows the early phase of star formation (SF). 
\end{abstract}


%% subject header.  Objects should be in the appropriate "individual"
%% headers (e.g. quasars: individual, stars: individual, etc.) with the
%% additional provision that the total number of headers, including each
%% individual object, not exceed six.  The \objectname{} macro, and its
%% alias \object{}, is used to mark each object.  The macro takes the object
%% name as its primary argument.  This name will appear in the paper
%% and serve as the link's anchor in the electronic edition if the name
%% is recognized by the data centers.  The macro also takes an optional
%% argument in parentheses in cases where the data center identification
%% differs from what is to be printed in the paper.



\section{Introduction}
\label{intro}

Star clusters are expected to form over a wide spectrum of masses that is later modified by the selective destruction of low-mass clusters \citep{fall77, gnedin97, vesperini97, vesperini98}. As  summarized in \citet{fall01}, star clusters are in general weakly bound and can be easily disrupted by a variety of mechanisms that operate on different time scales. On short ($t\la 10^7\, {\rm yr}$) and intermediate ($10^7 \la t \la {\rm few} \times 10^8\, {\rm yr}$) time scales, stellar evolution removes mass from star clusters by a combination of stellar winds, supernovae and other ejecta. On longer time scales ($t \ga {\rm few} \times 10^8\, {\rm yr}$) dynamical processes such as internal two-body relaxation, gravitational shocks, and dynamical friction become the main causes of mass loss. Mass loss through stellar evolution, in particular, seems to play a key role in shaping the mass function of the star cluster systems, from the initial power-law, that characterizes the mass function of the star forming regions, to the bell-shaped mass function of the globular clusters \citep[e.g. ][]{vesperini98, vesperini03}.

It is now believed that the fast disappearance of extremely young ($<10$ Myr) clusters, often known as ``infant mortality'' \citep{lada03}, is due to the rapid gas dispersal caused by stellar winds \citep[e.g.][]{whitworth79}, low-mass stars outflows \citep{matzner00}, and early-type star supernova explosions \citep{eggleston06}. 
The gas expulsion decreases the binding energy of the initial stellar system and, as a consequence, stars may suddenly have velocities that are higher than the local escape velocity \citep{tutukov78, hills80,  goodwin97}. The probability that a star cluster will survive the gas dispersal phase depends largely on the efficiency of SF,  a property that is still poorly quantified \citep{elmegreen07, price09}. 
Recently, \citet{Smith11}  have studied the effects of gas expulsion on sub-structured clusters formed under non-equilibrium initial conditions. They find that the initial spatial and kinematic distributions of the stars are far more important for cluster survival than the SF efficiency. They suggest that variations in cluster initial conditions, rather than SF efficiencies, determine whether clusters survive gas expulsion or not.
In view of these findings, in depth observational studies of the characteristics of young resolved star clusters are clearly needed. Measurements such as their PDMF, the presence of mass segregation, the stellar concentration and density distribution can provide much needed constraints for understanding the evolution and longevity of star cluster systems.

Because of its close proximity \citep[60.6 Kpc; ][]{hilditch05}, the SMC is uniquely suited for detailed investigations of the stellar content of regions of SF and young star clusters.  With HST it is possible to spatially resolve the densest star clusters and perform a quantitative and accurate census of the stellar content down to the sub-solar mass regime \citep[e.g.][]{sabbi08}.
As part of a project devoted to studying the properties of young and massive star clusters in the SMC (P.I. A. Nota, GO-10248), in this paper we present an in-depth study of the stellar content of the, so far,  poorly investigated star cluster NGC~376 ($\alpha_{J2000}=01^h 03^m 53.5^s,\, \delta_{J2000}=-72\degr 49' 27.0''$) based on HST imaging.
NGC~376 is one of the brightest, richest, and youngest (from 16 Myr, Chiosi et al. 2006, to 25 Myr, Piatti et al. 2007) clusters in the eastern extension of the SMC toward the LMC, known also as the SMC wing.  Previous studies suggest that the surface brightness profile of NGC~376 departs from a standard EFF model \citep{elson87} and that  it may be merging with another cluster \citep{carvalho08}.

The paper is organized as follows: in Section~2 we present the observations and the analysis of the data, while Section~3 is dedicated to the  description of the color-magnitude diagram (CMD) and of the stellar populations found in the region. In Section~4 we discuss the structural parameters of NGC~376 as derived from the surface brightness and the stellar density profiles. In Section~5 we derive the age of the cluster from isochrone fitting. We present the luminosity function (LF) and the PDMF in Section~6. The results are discussed in Section~7. 

\section{OBSERVATIONS AND DATA REDUCTION}

\subsection{The data}
\label{data}

We used the HST/ACS Wide Field Channel (WFC) to acquire deep images of NGC~376. The data were obtained in September 2004 as a part of a program devoted to study the characteristics of the youngest star clusters in the SMC (GO-10248; P.I. A.~Nota). We acquired four 450 sec long exposures (total integration time $=$ 1800 sec) in the F555W filter ($\sim$V), and four 490 sec long exposures (total integration time=1960 sec) in the F814W filter  \citep[$\sim$I -- see ][for a description of the ACS filters]{sirianni05}. Both F555W and F814W images were acquired following a four pointing dither pattern designed to better remove hot pixels and fill the gap between the two ACS/WFC $2048\times 2048$ CCDs. Dithering also allowed us to better sample the point-spread function (PSF) and, by averaging the flat-field errors and smoothing over the spatial variation of the detector response, to improve the photometric accuracy. To recover the photometric information for the brightest, and otherwise saturated, stars, we acquired two short exposures (3.0 sec) in both filters. A summary of the observations is given in Table~\ref{t:obs}.

\begin{deluxetable}{cccccc}
\tablecaption{{Journal of WFC/ACS observations}\label{t:obs}}
\tablehead{
\colhead{Image Name} & \colhead{Date \& Time of Observation} & \colhead{R.A.} & \colhead{Dec.} & \colhead{Filter} & \colhead{Exposure Time}}

\startdata
J92F08LZQ & 12/09/04  \, 13:52:45 & $01^{\rm h}03^{\rm m}53\fs8$ & $-72\arcdeg49\arcmin29\farcs6$ & F555W &  3.0 \\     
J92F08M0Q & 12/09/04  \, 13:55:06 & $01^h03^m53\fs8$ & $-72\arcdeg49\arcmin29\farcs6$ & F555W & 450.0 \\
J92F08M2Q & 12/09/04 \, 14:05:12 & $01^{\rm h}03^{\rm m}53\fs8$ & $-72\arcdeg49\arcmin29\farcs6$ & F555W & 450.0 \\
J92F08M4Q & 12/09/04 \, 14:15:18 & $01^{\rm h}03^{\rm m}53\fs8$ & $-72\arcdeg49\arcmin29\farcs6$ & F555W & 450.0 \\
J92F08M6Q & 12/09/04 \, 14:25:24 & $01^{\rm h}03^{\rm m}53\fs8$ & $-72\arcdeg49\arcmin29\farcs6$ & F555W & 450.0 \\
J92F08M9Q & 12/09/04 \, 14:41:01 & $01^{\rm h}03^{\rm m}53\fs8$ & $-72\arcdeg49\arcmin29\farcs6$ & F555W &   3.0 \\
J92F08MBQ & 12/09/04 \, 14:49:52 & $01^{\rm h}03^{\rm m}53\fs8$ & $-72\arcdeg49\arcmin29\farcs6$ & F814W &   3.0 \\
J92F08MCQ & 12/09/04 \, 14:52:13 & $01^{\rm h}03^{\rm m}53\fs8$ & $-72\arcdeg49\arcmin29\farcs6$ & F814W & 490.0 \\
J92F08MEQ & 12/09/04 \, 15:29:33 & $01^{\rm h}03^{\rm m}53\fs8$ & $-72\arcdeg49\arcmin29\farcs6$ & F814W & 490.0 \\
J92F08MGQ & 12/09/04 \, 15:40:19 & $01^{\rm h}03^{\rm m}53\fs8$ & $-72\arcdeg49\arcmin29\farcs6$ & F814W & 490.0 \\
J92F08MIQ & 12/09/04 \, 15:51:05 & $01^{\rm h}03^{\rm m}53\fs8$ & $-72\arcdeg49\arcmin29\farcs6$ & F814W & 490.0 \\
J92F08MLQ & 12/09/04 \, 16:07:22 & $01^{\rm h}03^{\rm m}53\fs8$ & $-72\arcdeg49\arcmin29\farcs6$ & F814W &   3.0 \\
\enddata
\end{deluxetable}

\begin{figure}
\epsscale{1.0}
\plotone{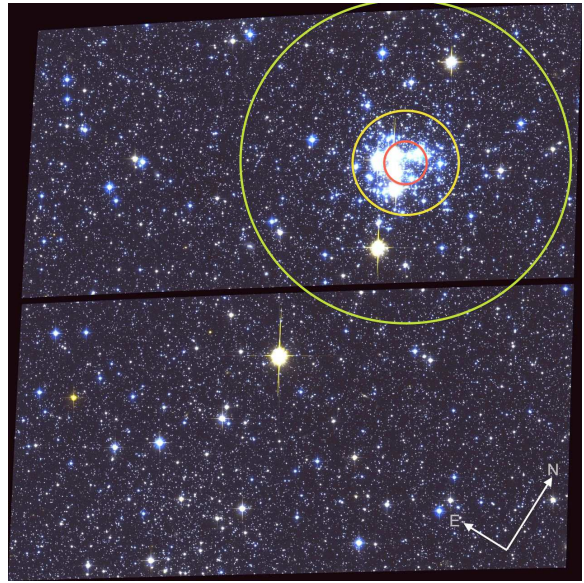}
\caption{\label{ima_376} ACS/WFC color-composite image of NGC~376. The image acquired through the filter F555W is in blue, while the F814W image is in red. North and East directions are indicated. The solid red circle defines the area within the core radius, the yellow solid circle marks the area within the tidal radius, and the green solid circle shows the extent of the cluster's tail.}
\end{figure}

All the images were acquired with a gain of $2 e^-\, {\rm ADU}^{-1}$, and the entire data set was processed through the standard Space Telescope Science Institute ACS calibration pipeline CALACS to subtract super-bias, and super-dark, and to apply the flat-field correction. Our ACS/WFC images cover an area of $200\arcsec \times 200\arcsec $ and have a pixel scale of $0.05\arcsec\, {\rm pixel}^{-1}$. Assuming a distance modulus $(m-M)_0=18.912\pm0.035$ \citep[corresponding to $\sim 60.6\, {\rm kpc}$,][]{hilditch05} the ACS/WFC field of view corresponds to a projected area of $\sim 50\times 50\, {\rm pc^2}$. A color composite ACS/WFC image of NGC~376 is shown in Figure~\ref{ima_376}.

\subsection{Photometric Reduction}
\label{photometry}

The photometric analysis was carried out directly on the pipeline-corrected images with extension {\sc \_flt}, using the program img2xym\_WFC.09X10 \citep{anderson06}, that was specifically designed to perform photometric analysis of the under-sampled ACS/WFC data. A library of empirical PSFs, that takes into account the PSF spatial variation due to the telescope optics and the variable charge diffusion in the CCD \citep{krist03}, is provided with the code. Temporal variations in the PSF, caused by changes in the telescope's focus induced by spacecraft breathing, are also taken into account by fitting in each image the library PSF to the brighter (${\rm S/N} > 100$) stars.

We used the routine described in \citet{anderson08} to further refine the photometry. The program divides all images in to regions of $25\times 25$ pixels and, region after region, simultaneously finds and fits the stars in all the exposures. For each star found, the code provides the average X and Y coordinates and F555W and F814W magnitudes, the r.m.s. of coordinates and magnitudes, as well as an estimate of the psf-fitting quality.

\begin{figure}
\epsscale{1.0}
\plotone{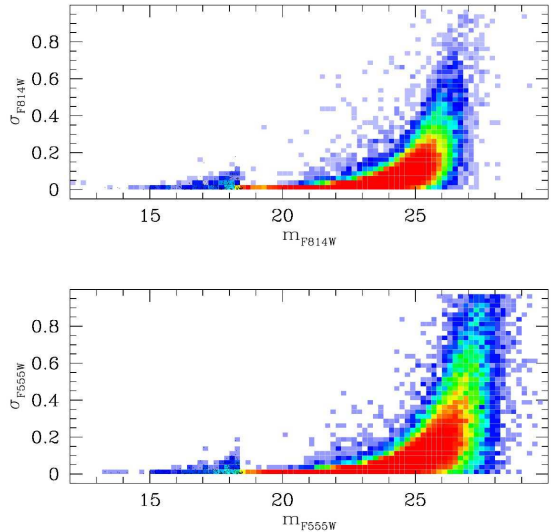}
\caption{\label{f:phot_err} Photometric r.m.s as a function of magnitude for stars that have been detected in at least three exposures after the selection in positional errors, in the F814W ({\it upper panel}) and the F555W ({\it lower panel}) filter respectively.}
\end{figure}

To remove as many spurious detections as possible from the final catalog, we retained only those stars that were identified in both the filters in at least three exposures with a positional error smaller than $<0.1$ pixel. Photometric r.m.s as a function of magnitude are shown in Figure~\ref{f:phot_err}.
%Figure~\ref{f:phot_err} shows the photometric r.m.s. as a function of magnitude for both the %filters, after the selections were applied. The final catalog contains more than 37,000 %stars.

To calibrate the final catalog in the Vegamag photometric system we combined the {\sc \_flt} images using the multidrizzle package \citep{koekemoer02} and performed aperture photometry for several isolated stars for both the F555W and F814W filters using DAOPHOT in IRAF\footnote{IRAF is distributed by the National Optical Astronomy Observatory, which is operated by AURA, Inc., under cooperative agreement with the National Science Foundation.}. These stars were calibrated into the Vegamag photometric system following the recipe by \citet{sirianni05} and then used to calibrate the photometric catalog.

\subsection{Artificial Star Tests: Photometric Errors and Completeness}
\label{art star}

\begin{figure}
\epsscale{1.0}
%\plotone{f3.ps}
\plotone{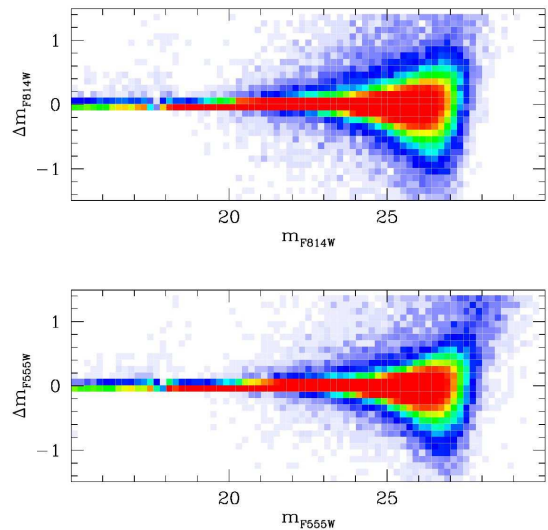}
\caption{\label{f:deltamag} Difference between the input and output magnitude of the artificial stars as a function of the input magnitude for both the F814W ({\it upper panel}) and the F555W ({\it lower panel}) filters.}
\end{figure}

\begin{figure}
\epsscale{1.0}
\plotone{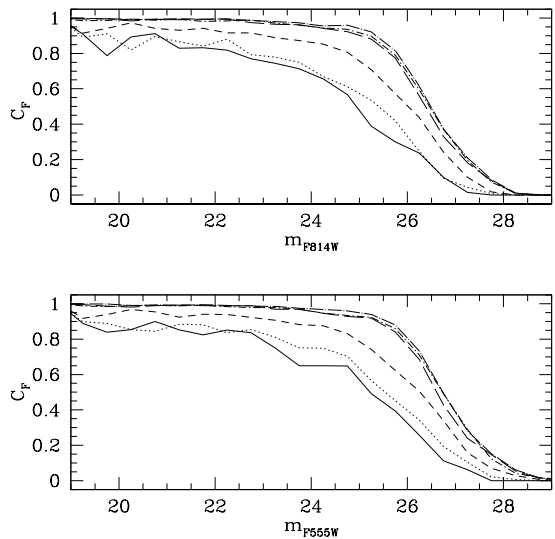}
\caption{\label{f:fc} The Completeness Factor $C_F$ as a function of the input magnitude as derived from the artificial star tests for both the F814W ({\it upper panel}) and the F555W ({\it lower panel}) filters as a function of the distance from the NGC~376 center. The solid lines represent the $C_F$s within $5\arcsec$ from the center, the $C_F$s in the annulus between $5\arcsec$ and $10\arcsec $ are represented by the dotted lines, the dashed lines are the $C_F$s between $10\arcsec$ and $15\arcsec$, the long-dashed lines, the dashed-dotted lines and and the long-dashed-dotted lines are the $C_F$s between $15\arcsec$ and $25\arcsec$, $25\arcsec$ and $35\arcsec$ and $45\arcsec$ and $55\arcsec$ respectively.}
\end{figure}

Artificial star tests are a standard procedure used to quantify the completeness of a photometric catalog, as well as to test the impact of crowding on the photometric accuracy. The tests are performed by inserting stars with known flux and position in the data set, and then repeating the photometric analysis using the same procedure applied to the real data. The difference between the input and output magnitude of the recovered artificial stars (Figure~\ref{f:deltamag}) provides an estimate of the photometric accuracy, while the photometric completeness, as a function of magnitude, is derived from the fraction of recovered artificial stars with respect to the simulated ones.  Figure~\ref{f:fc} shows the variation of the completeness factor ($C_F$) as a function of the distance from the center of NGC~376.

The same routines used in Section~\ref{photometry} to refine the photometry, can also perform artificial star tests. The program divides each image in to regions of $25\times 25$ pixels, and adds an artificial star in a region at a time. As during the photometric analysis, the program finds and fits all the sources in the region simultaneously in all the frames  \citep{anderson08}. This approach avoids the problem that artificial stars interfere with each other, artificially increasing the crowding of the image, and thus altering the inferred completeness. In total we simulated more than 1,000,000 stars in each of the F555W and F814W exposures. 

We considered an artificial star as recovered if:
\begin{itemize}
\item The input and output fluxes agree to within 0.75 mag;
\item The input and output positions agree to within 1 pixel;
\item The star is found in at least three F555W and three F814W exposures, with a positional error $< 0.1$ pixel.
\end{itemize}  

We find that the completeness of our sample varies as a function of the distance from the cluster (Figure~\ref{f:fc}). For example within the innermost $5\arcsec$ from NGC~376 the sample is 50\% complete to $m_{\rm F555W}=25.25$, between 5 and $10\arcsec$ the same completeness is at  $m_{\rm F555W}\simeq25.5$, and it goes down to $m_{\rm F555W}=26.25$ between 10 and $15\arcsec$. At $m_{\rm F555W}=24.75$ and $m_{\rm F814W}=24.25$ the sample is $>65\%$ complete over the entire field.

\section{COLOR-MAGNITUDE DIAGRAMS}
\label{cmd}

\begin{figure}
\epsscale{1.0}
\plotone{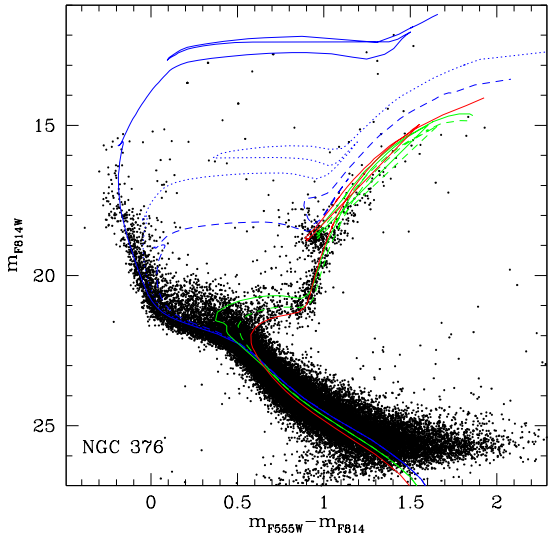}
\caption{\label{f:cmd} ACS/WFC CMD $m_{\rm F814W}$ vs. $m_{\rm F555W}-m_{\rm F814W}$ of NGC~376 with superimposed Padua isochrones of different ages and metallicities. The blue continuous isochrone corresponds to a 28 Myr old stellar population with metallicity Z$=$0.004, the dotted blue line corresponds to a Z$=$0.004 200 Myr old stellar population, and the dashed blue isochrone is for a Z$=$0.004 500 Myr old population. The green continuous and the long-dashed lines correspond to Z$=$0.002 isochrones for a 3 and 5 Gyr old stellar populations respectively. Finally the red line is a Z$=$0.001, 8 Gyr old isochrone.}
\end{figure}

The CMD $m_{\rm F814W}$ versus $m_{\rm F555W}-m_{\rm F814W}$ of all the stars that passed our selection criteria is shown in Figure~\ref{f:cmd}. At a distance of 60.6 kpc \citep{hilditch05} and assuming a metallicity Z=0.001 \citep{carrera08, parisi10}, Padua isochrones \citep{bertelli08, bertelli09} predict that the 13 Gyr main sequence turn-off (MSTO) is at $m_{\rm F814W}=21.88$. Our photometry therefore reaches $\ga 4$ magnitudes below the expected oldest MSTO. Padua isochrones \citep{bertelli08,bertelli09} for Y=0.24 and different ages and metallicities are superimposed on the CMD of Figure~\ref{f:cmd} for reference. To plot the isochrones we assumed a distance modulus of $(m-M)_0=18.9$, %\citep{hilditch05}
the Galactic extinction law, a reddening of $E(B-V)=0.08$ and $A_{F555W}=0.19$.

An inspection by eye of Figure~\ref{f:cmd} shows that, as found in several other SMC stellar fields \citep[i.e.][]{sirianni02, mccumber05, nota06, sabbi07, chiosi07, carlson07, noel07, sabbi09}, stellar populations of different ages coexist in the area. 

{\bf Old stars:} A stellar population older than $\sim 2\, {\rm Gyr}$ can be easily distinguished in the CMD. Evolutionary phases associated with this population are: 
\begin{itemize}
\item The well defined lower main sequence (MS), which extends from $m_{\rm F814W}\simeq 21.7$ down to $m_{\rm F814W}\la 26.4$; 
\item The broad ($20.3\la m_{\rm F814W}\la 21.6$) subgiant branch (SGB), visible in the color range $0.42\la m_{\rm F555W}-m_{\rm F814W}\la 0.90$. Comparison with the isochrones suggests that the majority of the stars in this evolutionary phase are $\sim 4-5\, {\rm Gyr}$ old;
\item The bright red giant branch (RGB), with the brightest star at $m_{\rm F814W}\simeq 14.9$ and $m_{\rm F555W}-m_{\rm F814W}\simeq 1.94$. The relatively small number of RGB stars does not allow us to establish if the brightest stars of the RGB correspond or not to the RGB-tip; 
\item The red clump (RC), at $m_{\rm F814W}\simeq 18.5$.
\end{itemize}
Asymptotic giant branch (AGB) stars are likely present in the CMD, but it is difficult to separate them from RGB stars.

{\bf Young and intermediate-age stars:} The bright ($15.0\la m_{\rm F814W} \la 21.0$) and blue ($m_{\rm F555W}-m_{\rm F814W}\la 0.2$) well-populated upper MS indicates that in this region the SMC was still forming stars less than $\sim 2\, {\rm Gyr}$ ago. Young low mass stars are likely present in the lower MS, however they cannot be distinguished from the older ($\ga 2\, {\rm Gyr}$) MS.
This finding is also supported by the presence of stars, above the RC, in the magnitude range $16.5\la m_{\rm F814W}\la 18.0$ and color range $0.56\la m_{\rm F555W}-m_{\rm F814W}\la 0.89$ which likely corresponds to the lower end of the blue edge of the blue loop and thus are younger than $\sim 500\, {\rm Myr}$.

Between $15.0\la m_{\rm F814W}\la 17.3$ and $0.0\la m_{\rm F555W}-m_{\rm F814W}\la 0.16$ there is a secondary sequence of objects that are likely Be stars. Their presence in the CMD suggests that this region was still forming stars $\sim 30\, {\rm Myr}$ ago \citep{keller00}. This is also confirmed by the few red super giant stars (RSGs) brighter than $m_{\rm F814W}>13.1$ and redder than $m_{\rm F555W}-m_{\rm F814W}>1.24$. 

In two other SMC star clusters \citep[namely NGC~346 and NGC~602, ][]{nota06, sabbi07, carlson07, cignoni09, cignoni11} a similar observational set-up allowed us to find rich populations of pre-main sequence stars on the right side of the lower MS. The absence of pre-main sequence stars in Figure~\ref{f:cmd} suggests that in this region of the SMC very few stars, if any, were formed in the last $\sim 15\, {\rm Myr}$. 

Finally we do not find any clear evidence of horizontal branch (HB) stars in the CMD of Figure~\ref{f:cmd}, which are considered unequivocal indicators of a metal poor stellar population, older than $\sim 10\, {\rm Gyr}$, however the presence of an old-metal poor population in the SMC is confirmed by the finding of several RR-Lyrae over the entire SMC \citep[e.g.][]{soszynski02}. The paucity of HB stars has been confirmed by other photometric studies of the SMC \citep[i.e.][]{noel07, sabbi09}, and this apparent dichotomy between the presence of RR-Lyrae and the absence of extended HB has been interpreted by several authors \citep[i.e.][]{chiosi07, noel10} as an indication that the SMC formed very few stars in the first 2-3 Gyr .

\section{NGC~376 STRUCTURAL PARAMETERS}
\label{param}

Once formed, the evolution of  a star cluster is affected by a continuous  mass loss caused by gas expulsion, low-mass star evaporation and stellar evolution. 
Surface brightness and stellar density profiles are commonly used to probe the dynamical status of a star cluster. The first step in building a radial profile is to determine the center of the stellar population. Since stellar luminosities are not always proportional to stellar masses,  the center of luminosity ($C_{lum}$) can differ significantly from the center of gravity ($C_{grav}$) of a cluster. 

In the F555W image the $C_{lum}$ is at $\alpha=01{\rm ^h}03^{\rm m}53.9^{\rm s}\pm 0.01^{\rm s};\, \delta=-72\degr49\arcmin 34.0\arcsec \pm 0.3\arcsec$. We exploited the high spatial resolution of our photometric catalog to determine also the $C_{grav}$ of NGC~376. To find the $C_{grav}$ we iteratively averaged the stellar X and Y coordinates \citep{montegriffo95}. In order to take into account effects due to crowding and incompleteness, we computed the $C_{grav} $ using three different magnitude thresholds ($m_{\rm F555W}\le 22$, 23, and 24). Finally to test the impact of the SMC field on our measurement, we derived the $C_{grav}$ both using all the stars, and using only stars bluer than $m_{\rm F555W}-m_{\rm F814W}\le 0.5$. Our final best estimate of the NGC~376 $C_{grav}$ is  $\alpha=01^h 03^m 52.98^s \pm 0.01;\,  \delta=-72\degr 49\arcmin 32.5\arcsec \pm 0.15\arcsec$.

\subsection{The surface brightness profile}
\label{brightness}

NGC~376 has been recently observed in the V band with the ESO Danish 1.54 m telescope, in La Silla \citep{carvalho08}. The derived surface brightness profile shows a central peak, and an external ``bump''  that cannot be reproduced with a standard EFF model \citep{elson87}. 
 \citet{carvalho08} interpreted these anomalies as the result of a recent, or even still ongoing, merger with another star cluster.

\begin{figure}
\epsscale{1.0}
\plotone{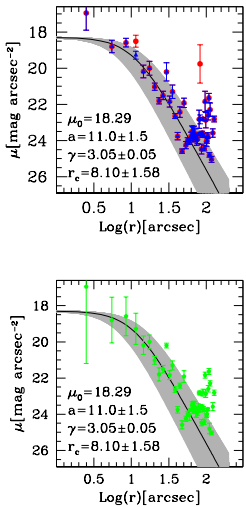}
\caption{\label{f:eff} NGC~376 surface brightness profile with the EFF models superimposed. The black line represents the best fit to our data, while the gray region represents the uncertainties to the fit. In the Top Panel the average surface brightness of the field of the SMC has been measured from SF10. Red dots are for all the stars brighter than $m_{\rm F555W}\le 24$, while blue triangles are derived considering only stars between $14< m_{\rm F555W}\le 24$. In the Bottom Panel the surface brightness of the SMC field has been estimated from the stars at a distance larger than $120\arcsec$ from the cluster.}
\end{figure}

To measure the surface brightness profile of NGC~376 (Figure~\ref{f:eff}) we divided our catalog into equally spaced annuli centered on the $C_{lum}$. Because of the cluster location in the upper right corner of the image (Figure~\ref{ima_376}) only the annuli in the range $0\arcsec - 50\arcsec$ are fully imaged, while at larger radii only a portion of each annulus falls in our field of view. For this reason we divided each annulus into 12 sectors, and for each annulus we considered only those sectors that fully lie in our field of view. 

Figure~\ref{f:fc} shows that for stars brighter than  $m_{\rm F555W}\la 24$ our catalog is more than 60\% complete at each distance from the cluster center. We, therefore, used only stars brighter than this threshold to measure the surface brightness profile. Figure~\ref{f:fc} shows also that, because of the increasing crowding, the completeness correction changes as a function of the distance from NGC~376. To apply the right correction to each point of the profile, in each annulus we divided the stars of each sector into bins of magnitude, and to each magnitude bin we applied the completeness factor $C_F$ as measured at that distance from the cluster center. 

We used the recipe from \citet{sirianni05} to convert the F555W magnitudes into fluxes. The brightness of each sector was obtained by summing the flux of all the stars brighter than $m_{\rm F555W}\le 24$. The flux of each annulus was obtained from the average of its sectors.

The well defined RGB and SGB in Figure~\ref{f:cmd} indicate that a considerable number of stars in our catalog belong to the field of the SMC. To statistically remove the contribution of the SMC from the surface brightness profile, we measured the average surface brightness in SF10, a field in the wing of the SMC ($\alpha=01^h 08^m 39^s; \delta=-72\degr 58\arcmin 45\arcsec$) that was observed in January 2006 with a similar observational strategy with the ACS/WFC in the F555W and F814W filters (GO-10396, P.I. J.S. Gallagher), as part of a project devoted to studying the star formation history (SFH) of the field of the SMC \citep{sabbi09} as well as to characterizing the properties of intermediate and old star clusters \citep{glatt08a, glatt08b}. 

The surface brightness profile of NGC~376, after the subtraction of the stars SMC field, is shown in the top panel of Figure~\ref{f:eff} (red dots).   To fit this profile we used an EFF model expressed in magnitude per surface area, rather than luminosity, using the formula 
$$l\mu(r)=l\mu_0(r)+1.25\gamma Log(1+r^2/a^2)$$
where $l\mu_0$ is the central surface brightness \citep[corresponding to $-2.5Log\mu_0(r)$ in the original formula presented in][]{elson87}, $\gamma$  is the slope of the power-law, and $a$ is a dimensionless parameter that is related to the core radius $r_c$ of the King profile by $r_c = a(2^{2/\gamma}-1)^{1/2} (= 8.10\pm 1.58$ arcsec). The surface brightness profile is quite irregular, and shows an evident central  peak that cannot be reproduced by our best fit. In agreement with the findings of  \citet{carvalho08}, beyond $\sim 60\arcsec$ there is a bump in the counts that exceeds the EFF profile. 

To verify that the irregularities found in the surface brightness profile are not caused by few bright foreground stars nor by local variations in the field of the SMC we repeated the analysis by first selecting only the stars fainter than $m_{\rm F555W}>14.0$ (blue triangles in the top panel of Figure~\ref{f:eff}), and then by measuring the average surface brightness at a distance larger than $120\arcsec$ from NGC~376 $C_{lum}$ (Figure~\ref{f:eff} -- bottom panel).

\subsection{The stellar density profile}
\label{density}

\begin{figure}
\epsscale{0.5}
%\plotone{f7.ps}
\plotone{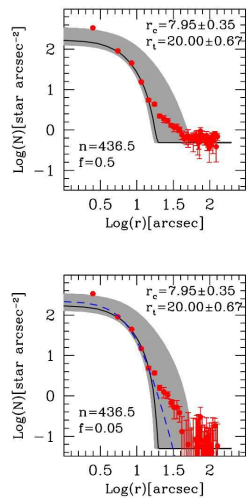}
\caption{\label{f:king} NGC~376 stellar density profile with superimposed the King profile (black line) in better agreement with the data. The grey region represents the uncertainty of the fit. The local average SMC stellar density has been estimated both from SF10 (Top Panel) and at a distance larger than $120\arcsec$ from the cluster (Bottom Panel). The dashed blue line in the bottom panel is EFF model that best fit the surface density profile shown in Fig.~\ref{f:eff}, scaled to the stellar density.  The inconsistency of the density profile with both the King's and EFF models is apparent. }
\end{figure}

Surface density profiles are dominated by the light from the brightest stars, but low mass stars play a dominat role in the dynamics of a star cluster. To better constrain NGC~376 dynamics, and taking advantage of the high spatial resolution of the ACS data, we also analyzed the stellar density profile of the cluster. Following our adopted procedure for the surface brightness profile, the stellar density profile was  obtained by dividing our sample in equally spaced annuli centered, in this case, on the $C_{grav}$. Each annulus was divided into 12 sectors to take into account the fact that, at a distance larger than $r>50\arcsec$, part of the annuli fall outside the image. In our analysis we considered only those sectors that fully lie in our field of view.  In each sector we divided the stars brighter than $m_{\rm F555W}\la 24$ into bins of magnitude and then we applied the completeness factor $C_F$, as measured at that distance from the cluster center,  to each magnitude bin. We obtained the stellar density of each sector by dividing the number of stars by the area of the sector. The average stellar density of the sectors in an annulus was used as the stellar density of that annulus. To estimate the density in the field of SMC we used both the average stellar density of SF10 (Figure~\ref{f:king} -- Top Panel) and the average stellar density at a distance larger than $120\arcsec$ from the cluster $C_{grav}$ (Figure~\ref{f:king} -- Bottom Panel). The radial density profile of NGC~376, after the subtraction of the SMC background, is shown in Figure~\ref{f:king}. 

King models  \citep{king66} can be used to infer the dynamical status of a star cluster. King profiles are usually described by a core radius $r_c$, that corresponds to the distance from the center where the projected stellar density falls to 0.5013, and a tidal radius $r_t$, that indicates where the potential well of the cluster equals zero. The ratio of the two radii ($r_t$ to  $r_c$) defines the concentration of the star cluster ($c={\rm log}_{10}(r_t/r_c))$ and it can be used to establish if a stellar system is in virial equilibrium or not \citep{meylan97}:
\begin{itemize}
\item Models with $c$ between 0.75 and 1.75 fit relaxed star clusters very well.
\item Models with $c\ga 2.2$ fit core collapsed globular clusters. 
\item Stellar systems described by a King model with $c<0.7$ are not in virial equilibrium.
\end{itemize}

The King model that best fits our data ($r_c=7.95 \pm 0.35; r_t=20.0 \pm 0.67$) provides a quite low concentration ($c\simeq 0.4$), suggesting that the cluster is not in virial equilibrium. To facilitate a more direct comparison between the EFF and King models, in Figure~\ref{f:king} (Bottom Panel) we plotted also the best fit EFF model, derived in Section~\ref{brightness}. This comparison shows that a single King (or EFF) model cannot fit the entire stellar density profile. Similar to what we found for the surface brightness, our best fit underestimates the stellar density  in the innermost $\sim 3-4\arcsec$, and cannot reproduce the tail of counts that continues well beyond (up to $\sim 60\arcsec$) the tidal radius ($r_t=20\pm 0.67\arcsec$) estimated from the best fit of the stellar density profile. 

Dynamical simulations \citep{kuepper10} using the velocity dispersion profile show that  the number of potential escapers from the star cluster increases and, sometimes even dominates the surface density profile,
for radii larger than $\sim$ half the Jacobi radius (often approximated to the $r_t$ of the King model). In this context the comparison between the surface brightness and the stellar density profiles seems to suggest that a considerable fraction of the stars that were formed in NGC~376 are not gravitationally bound to the cluster any more. 

\section{THE AGE}
\label{age}

\begin{figure}
\epsscale{1.0}
\plotone{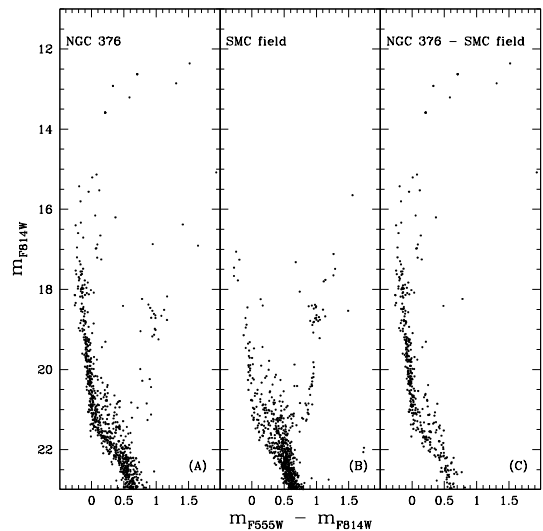}
\caption{\label{f:clean} CMDs $m_{\rm F814W}$ vs. $m_{\rm F555W}-m_{\rm F814W}$ of the stars within the tidal radius of NGC~376 ({\it panel (A)}), the SMC field, as observed at $r>120\arcsec$ from the center on NGC~376 and normalized to the area of the cluster ({\it panel (B)}). {\it Panel (C)} shows the CMD of NGC~376 after the decontamination from the SMC field.}
\end{figure}

Figure~\ref{f:clean} -panel (A) shows the CMD of all the stars brighter than $m_{\rm F814W}<23$ found within the NGC~376 tidal radius. The most evident feature in the CMD is the tight blue ($m_{\rm F555W}-m_{\rm F814W}\simeq 0.0$) MS, likely populated by NGC~376 stars. However the presence of SGB and RGB stars redder than $m_{\rm F555W}-m_{\rm F814W}>0.4$ indicates that even inside the cluster the contribution of stars belonging to the field of the SMC is not negligible.  To characterize the field of the SMC around NGC~376 we selected stars brighter than $m_{\rm F814W}<23$ that are at a distance $r>120\arcsec$ from the $C_{grav}$ of the cluster. This selection allowed us to avoid the population of ``potential escapers'' that likely dominates between $\sim 20 - 60\arcsec$ (Figure~\ref{f:king}). The CMD of the field, normalized to the area covered by NGC~376, is shown in Figure~\ref{f:clean} - panel (B), while the CMD of NGC~376 after the subtraction of the SMC field is shown in Figure~\ref{f:clean} - panel (C). 

\begin{figure}
\epsscale{1.0}
\plotone{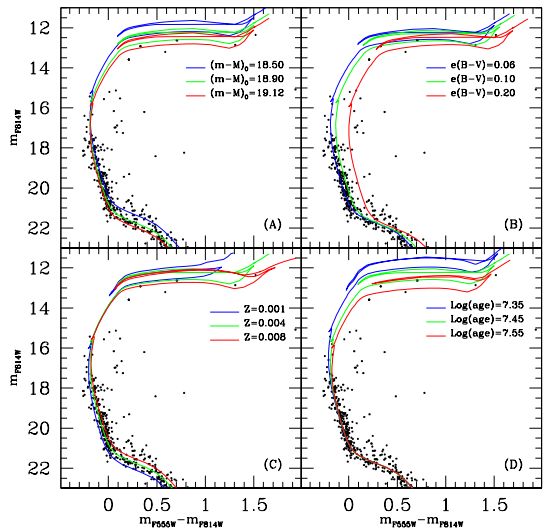}
\caption{\label{f:age_error} CMD $m_{\rm F814W}$ vs. $m_{\rm F555W}-m_{\rm F814W}$ of NGC~376, after the subtraction of the SMC field, with a 28 Myr Padua isochrone computed for Z$=$0.004, assuming an E(B-V)$=0.06$ and different distance moduli ({\it panel (A)}); a 28 Myr Padua
isochrone computed for Z$=$0.004, assuming a distance modulus (m-M)$_0=18.9$ but different values of reddening ({\it panel (B)}; 28 Myr Padua isochrones computed for different metallicities, assuming a distance modulus (m-M)$_0=18.9$ and an E(B-V)$=0.06$ ({\it panel (C)}). {\it Panel (D)} shows Z$=$0.004 Padua isochrones in the interval of time between 32.4 and 35.5 Myr assuming a distance modulus (m-M)$_0=18.9$ and an E(B-V)$=0.06$.}
\end{figure}

We used Padua isochrones  \citep{bertelli08, bertelli09} computed for Y=0.24 and various Z to infer the age of NGC~376 from the $m_{\rm F814W}$ vs. $m_{\rm F555W} - m_{\rm F814W}$ CMD. To reproduce the tight MS of NGC~376 we started from the literature values for distance, reddening and metallicity, and then we varied one parameter at the time. The effects of these changes are shown Figure~\ref{f:age_error}. 

We chose for the distance modulus of the SMC the value derived by \citet{hilditch05} from the analysis of more than 50 eclipsing OB star binary systems $(m-M)_0=18.912\pm 0.035$. The binaries are scattered over the whole galaxy, and therefore are likely representative of the galaxy mean distance. It has to be noted that several authors \citep[e.g.][]{mathewson88, hatzidimitriou93, crowl01, lah05, glatt08a} found that the depth of the SMC can be up to 20 kpc. In the sample of clusters analyzed by \citet{glatt08a}, for example, the closest object has a distance modulus $(m-M)_0=18.5$, while the distance modulus of the farthest system is $(m-M)_0=19.12$. Panel (A) of Figure~\ref{f:age_error} shows that among this range of values our data are well fitted by the average distance modulus $(m-M)_0=18.9$. Similarly panel (B) of Figure~\ref{f:age_error} shows that the average SMC reddening value E(B-V)=0.08 \citep{zaritsky02} matches our data well. We also considered isochrones for three different metallicity values (Z=0.001, Z=0.004 and Z=0.008 -- Panel (C) of Figure~\ref{f:age_error}) and found that a metallicity of 1/5 solar is the most adequate to reproduce both the colors and magnitudes of NGC~376 MS stars from $m_{\rm F814W}\sim 16$ down to $m_{\rm F814W}\sim 23$.
In summary, assuming a distance modulus $(m-M)_0=18.9$, a reddening ${\rm E(B-V)}=0.08$ and a metallicity Z$=$0.004, Padua isochrones indicate that NGC~376 has an age of $28\pm 7$ Myr (Figure~\ref{f:age_error} -- panel (D)), in good agreement with the results by \citet{piatti07} and with the finding that NGC~376 is hosting several Be stars. %The tight MS of NGC~376 suggests that in this cluster all the stars were likely formed over a short interval of time, making the hypothesis that NGC~376 has been perturbed by recent, or by an ongoing, merger with another cluster \citep{carvalho08} quite unlikely.

\section{PRESENT DAY LUMINOSITY AND MASS FUNCTIONS}
\label{mf}

\begin{figure}
\epsscale{1.0}
\plotone{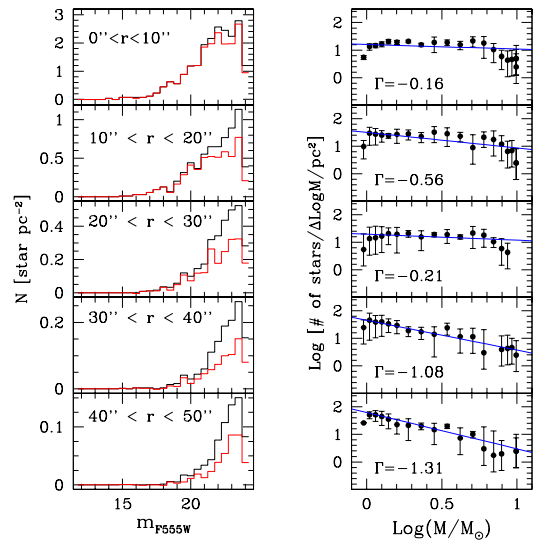}
\caption{\label{f:mf} LFs ({\it left column}) and MFs ({\it right column}) for different annuli ($0\arcsec-10\arcsec; 10\arcsec-20\arcsec; 20\arcsec-30\arcsec; 30\arcsec-40\arcsec; 40\arcsec-50\arcsec$) around NGC~376 $C_{grav}$. In the left column black histograms are the LFs, corrected for completeness, before the SMC background subtraction, and red histograms are the LFs after the subtraction of the SMC field. The slope $\Gamma$ of the MF, as derived from the weighted least squares fit to the data, is shown in each of the right panels. }
\end{figure}

We now examine the stellar PDMF of NGC~376 and derive the total mass of the cluster.
We used a procedure analogous to that described in Section~\ref{param} to measure the LF of NGC~376 down to $m_{\rm F555W}=24$ and to apply a correction for completeness as a function of the distance from the cluster. To remove the contribution of the SMC stars from the LF we used the LF of the stars found at a distance larger than $r> 120\arcsec$. The catalog was corrected for completeness and normalized for the area. The LFs in the F555W band as a function of the distance from NGC~376 $C_{grav}$ are shown in Figure~\ref{f:mf} (left column) before (black histogram) and after (red histogram) the subtraction of the SMC background. 

To convert the observed LFs in to PDMFs we used the mass--luminosity (ML) relation derived from the 28 Myr old, Z=0.004, Y=0.24  Padua isochrone \citep{bertelli08, bertelli09}. Among the available parametrizations, we chose the one proposed by \citet{scalo86}, in which the MF is characterized by the logarithmic derivative $\Gamma=d {\rm log} \xi({\rm log}(m))/d$, where $\xi({\rm log}(m))$ is the MF and $\Gamma$ is its slope. In this parametrization, the slope of the solar neighborhood initial mass function derived by \citet{salpeter} is $\Gamma=-1.35$. 

We calculated the derivative of the ML relation in the mass range between 1--10 M$_\odot$ using a spline interpolation. The PDMFs derived in this way for different radii are shown in Figure~\ref{f:mf} (right column). The derived slopes are affected by several uncertainties, such as the assumptions on distance, amount of extinction, and age as well as residual contamination from the field. A major source of uncertainty for NGC~376 likely comes from the unresolved binary systems. In a dynamically mass segregated cluster, the most massive binaries should be confined in the center with respect to lighter systems, further steepening the distribution. \citet{sagar91} estimated that if each star in the mass range 2--14 M$_\odot$ has one companion, the average IMF slope derived for five young clusters in the LMC would significantly steepen. More detailed modeling by \citet{weidner09}, however, suggests that binaries have little effect on the measured slope of the MF.

A weighted least-mean square fit of the data indicates that the PDMF remains flat ($\langle \Gamma \rangle=-0.31\pm 0.27$) over the entire cluster extent, and it becomes close to the value found by Salpeter only in the cluster's tail. As for other young star clusters  \citep{degrijs02, sirianni02, stolte02, gouliermis04, sabbi08} the steepening of the PDMF with distance from the center is caused by a lack of massive stars rather than an excess of low-mass stars in the outskirts. The paucity of massive stars in the halo of a star cluster is normally interpreted as a signature of mass segregation. Some of the known mass-segregated clusters are so young that mass-segregation has to be primordial \citep{hillenbrand97, hillenbrand98, sabbi08}. This does not seam to be the case for NGC~376: from the PDMF and the King profile we find that the stellar mass within the tidal radius $r_t$ is $M_{\rm TOT_{rt}}\simeq 7\times 10^4\, {\rm M_\odot}$, and, considering also the stars in the cluster's tail ($r\la 60\arcsec$), we find a total mass $M_{\rm TOT}\simeq 6\times 10^5\, {\rm M_\odot}$. It thus appears that only $\sim 10\%$ of the present day cluster mass is within the tidal radius. We can substitute the cluster virial radius with the half mass radius ($r_{vir}\approx r_{hm}$), in which case (considering only the stars within the tidal radius) the dynamical time scale ($T_{\rm dyn}$) necessary for NGC~376 to reach dynamical equilibrium is $T_{\rm dyn}\simeq 2.6\times 10^5$ yr , and, if we consider the mass within $60\arcsec$ from the $C_{grav}$, it rises to $T_{\rm dyn}\simeq 7.8\times 10^5$ yr , pointing toward a dynamical origin for the mass segregation observed in NGC~376. The PDMF also suggests that even within the tidal radius stars are weekly bound since at the tidal radius (which at the distance of the SMC correspond to $\sim5.9$ pc) the escape velocity is $\sim 3\, {\rm km/s}$. These findings suggest that the NGC~376  is likely evaporating into the field of the SMC. The properties of NGC~376 derived in this paper are summarized in Table~\ref{t:properties}.

\begin{deluxetable}{lr}
\tablecaption{{Properties of NGC~376, as derived in this paper}\label{t:properties}}
\tablehead{}
%\colhead{Image Name} & \colhead{Date \& Time of Observation} & \colhead{R.A.} & \colhead{Dec.} & \colhead{Filter} & \colhead{Exposure Time}}

\startdata
NGC~376 core radius from King's model					& $r_c = 7.95 \arcsec \pm 0.35\arcsec \simeq 2.34 \pm 0.10\, {\rm pc}$  \\
NGC~376 tidal radius from King's model					& $r_t=20.00\arcsec \pm 0.67\arcsec \simeq 5.88 \pm 0.20\, {\rm pc}$ \\
NGC~376 core radius from EFF model	 				& $r_c=8.10\arcsec \pm 1.58\arcsec \simeq 2.40 \pm 0.46\, {\rm pc}$ \\
PDMF slope within $20\arcsec$ from the center			& $\Gamma=-0.31\pm 0.27$ \\
PDMF slope between 40 and $60\arcsec$		& $\Gamma=-1.20\pm0.11$  \\
NGC~376 total mass								& $M{\rm TOT_{rt}}\simeq 7\times 10^4 {\rm M}_\odot$ \\
Total mass within $60\arcsec$ 							& $M{\rm TOT}\simeq 6\times 10^5 {\rm M}_\odot$ \\
NGC~376 dynamical time								& $T_{\rm dyn}\simeq 2.6\times 10^5\, {\rm yr}$\\
NGC~376 escape velocity							& $\sim 3\, {\rm km/s} $ \\
Assumed distance modulus							& $(m-M)_0=18.90$ \\ 
\enddata
\end{deluxetable}

\section{DISCUSSION AND CONCLUSIONS}
%\subsection{About the destiny of NGC~376}

We analyzed deep ACS/WFC images of the young star cluster NGC~376 in the filters F555W and F814W, as part of a project devoted to study the formation and evolution of young star clusters in the SMC (P.I. A. Nota; GO--10248). Using Padua isochrones \citep{bertelli08, bertelli09} we found that NGC~376 is $28\pm 7$ Myr old, in agreement with the previous study by \citet{piatti07}. 

Previous analysis of the cluster surface brightness profile \citep{carvalho08} showed that it is quite irregular, with a spike in the center and an extended bump in the outer regions. We took advantage of the high spatial resolution of our dataset to repeat the analysis of the surface brightness profile and to measure, for the first time, the cluster stellar density distribution down to $m_{\rm F555W}=24$, which, at the distance of the SMC, for a 28 Myr old stellar population of Z=0.004 metallicity, corresponds to $\simeq0.96\, M_\odot$. We used an EFF model  \citep{elson87} to fit the surface brightness and a King model \citep{king66} to reproduce the stellar density profile. In both the cases, the best fits underestimate the central peak. Furthermore the stellar density profile shows a tail ($r_{\rm tail}\simeq 60\arcsec \simeq 17\, {\rm pc}$) that extends well beyond  the tidal radius ($r_t = 20\arcsec \simeq 5.9\, {\rm pc}$) derived from the best fit King model. Our analysis indicates that NGC~376 has a low concentration ($c=0.4$), typical of clusters that are not in virial equilibrium \citep{meylan97}. 

The anomalous surface brightness profile of NGC~376 was interpreted by \citet{carvalho08} as the result of a recent, or even a still ongoing, merger. However, no obvious interacting candidate, such as another star cluster was found in the vicinity of NGC~376, and the tight MS of NGC~376 suggests that all its stars were formed in a relatively short interval of time. Similarly we did not find any know giant molecular cloud (GMC) within $\sim 30\arcmin$, which at the distance of the SMC corresponds to a projected distance of $\sim530\, {\rm pc}$ \citep{mizuno01}. These circumstances seem to point towards a different source of heating.  \citet{bastian06} reported that, in some nearby starburst dwarf galaxies, super star clusters younger than 60 Myr show a significant deviation from King and EFF profiles. From the comparison between the observed luminosity profiles and dynamical simulations, \citet{goodwin06} proposed that these clusters are far from virial equilibrium, because of the likely expulsion of the residual gas after the process of SF. We therefore suggest, as an alternative to the merger hypothesis, that NGC~376 was perturbed by the rapid expulsion of gas that followed the initial phase of SF. Because of the drop in the initial binding energy of NGC~376, a large fraction of stars now have velocities higher than the escape velocity and are no longer  gravitationally bound to the cluster.

We studied how the PDMF varies as a function of the distance from the center of NGC 376. We found that the PDMF remains flat ($\langle \Gamma\rangle=-0.31\pm 0.27$) over the entire cluster, and it becomes close to the value derived by \citet{salpeter} for the IMF in the solar neighborhood only in the clusters's tail beyond the tidal radius.  If we assume that the stars found in the tail of stellar density profile between $\sim20 $ and $\sim60\arcsec$ belong to NGC~376, then almost 90\% of total stellar mass of the cluster is likely not bound to the cluster.
%in the form of low mass stars.. 
%As already found in other young star clusters (Hillenbrand 1997; Hillenbrand \& Hartmann 1998; Sirianni et al. 2002) the increase in the MF slope with the distance from the center is due to the lack of massive stars at the periphery rather than an excess of low-mass objects there. The ratio between the projected stellar density of the massive stars in the core and in the tail is 10 times higher than the ratio of the low mass star density in the same regions, suggesting 

Several authors \citep{zhan99, chandar06, degrijis06, degrijs08, gieles08, chandar10a, chandar10b} have recently analyzed the properties of the young clusters (less than few $10^8$ Myr) in both the Magellanic Clouds with the aim of understanding the disruption mechanisms of star clusters and how long they can survive. However, different groups derived contradicting conclusions, even when using the same data, with some favoring the infant mortality scenario \citep{carpenter00, lada03, withmore05}, that predicts that more than $80-90\%$ of the clusters will be disrupted in a short, nearly mass independent interval of time \citep[e.g. ][]{chandar06,chandar10b}, and others favoring a less efficient disrupting mechanism \citep[e.g. ][]{degrijis06,degrijs08}. 

From the King profile of NGC~376 and its PDMF we estimate that the escape velocity from the cluster at the tidal radius is $\sim 3\, {\rm km/s}$, suggesting that even within the tidal radius the stars are probably weakly bound, and that the cluster is likely evaporating. However, additional information on the kinematics of NGC~376 is necessary to confirm the fate of NGC~376. The fact that a cluster as massive as $M_{\rm TOT}\simeq 6\times 10^5\, {\rm M_\odot}$ can lose $\sim 90\%$ of its mass in less then $\sim 20\, {\rm Myr}$ seems to favor a fast disruptive mechanism. A systematic study of the dynamical and kinematic properties of resolved clusters younger than $10^8$ Myr can provide independent diagnostics to probe the time-scale of cluster disruption.

\acknowledgments
We thanks the referee for many constructive suggestions.
We are grateful to J. Anderson for the software support, and to N. Panagia, L.R. Bedin, and A. Bellini for useful discussions. M.C. and M.T. acknowledge financial support through contracts ASI-INAF-I/016/07/0 and PRIN-INAF 2008.


\begin{thebibliography}{}

\bibitem[Anderson \& King(2006)]{anderson06}
Anderson, J., \& King, I.R. 2006, PSFs, Photometry and Astronomy for the ACS/WFC (Instrum. Sci. Rep. ACS 2006-01; Baltimore, MD:STScI)

\bibitem[Anderson et al.(2008)]{anderson08}
Anderson, J., et al. 2008, \aj, 135, 2055


\bibitem[Bastian et al.(2005)]{bastian05}
Bastian, N., Gieles, M., Lamers, H.J.G.L.M., Scheepmaker, R.A., de Grijs, R. 2005, A\&A, 431, 905

\bibitem[Bastian \& Goodwin(2006)]{bastian06}
Bastian, N., \& Goodwin, S.P. 2006, MNRAS, 369, L9

\bibitem[Bastian et al(2006)]{bastian06b}
Bastian, N., Saglia, R.P., Goudfrooij, P., Kissler-Patig, M., Maraston, C., Schweizer, F., Zoccali, M. 2006, A\&A, 448, 881

\bibitem[Bastian et al.(2009)]{bastian09}
Bastian, N., Gieles, M., Ercolano, B, Guthermuth, R. 2009, MNRAS, 392, 868

\bibitem[Bertelli et al.(2008)]{bertelli08}
Bertelli, G., Girardi, L., Marigo, P., Nasi, E. 2008, A\&A, 484, 815

\bibitem[Bertelli et al.(2009)]{bertelli09}
Bertelli, G., Nasi, E., Girardi, L., Marigo, P. 2009, A\&A, 508, 355

%\bibitem[Binney \& Tremaine(1987)]{binney87}
%Binney, J., \& Tremaine, S. 1987, ``Galactic DynamicsÕÕ, Pricenton University Press, %Princeton, New Jersey

\bibitem[Carlson et al.(2007)]{carlson07}
Carlson, L.R., Sabbi, E., Sirianni, M., Hora, J.L., Nota, A., Meixner, M., Gallagher, J.S., Oey, M.S., Pasquali, A., Smith, L.J., Tosi, M., Walterbos, R. 2007, \apj, 665, L109

\bibitem[Carpenter(2000)]{carpenter00}
Carpenter, J.M. 2000, \aj, 120, 3139

\bibitem[Carrera et al.(2008)]{carrera08}
Carrera, R., Gallart, C., Aparicio, A., Costa, E., M\'endez, R.A., \& No$\ddot{e}$l, N.E.D. 2008, \aj, 136, 1039

\bibitem[Carvalho et al.(2008)]{carvalho08}
Carvalho, L., Saurin, T.A., Bica, E., Bonatto, C., Schmidt, A.A. 2008, A\&A, 485, 71

\bibitem[Chandar et al.(2006)]{chandar06}
Chandar, R., Fall, S.M., \& Whitmore, B.C. 2006, \apj, 650, L111%%

\bibitem[Chandar et al.(2010a)]{chandar10a}
Chandar, R., Fall, M.S., Whitmore, B.C. 2010a, \aj, 139, 545


\bibitem[Chandar et al.(2010b)]{chandar10b}
Chandar, R., Fall, S.M., Whitmore, B.C. 2010b, \apj, 711, 1263

\bibitem[Chiosi et al.(2006)]{chiosi06}
Chiosi, E., Vallenari, A., Held, E.V., Rizzi, L., Moretti, A. 2006, A\&A, 452, 179

\bibitem[Chiosi \& Vallenari(2007)]{chiosi07}
Chiosi, E., Vallenari, A. 2007, A\&A, 466, 165

\bibitem[Cignoni et al.(2011)]{cignoni11}
Cignoni, M., Tosi, M., Sabbi, E., Nota, A., Gallagher, J.S. 2011, AJ, 141, 31

\bibitem[Cignoni et al.(2009)]{cignoni09}
Cignoni, M., et al. 2009, \aj, 137, 3668

\bibitem[Crowl, et al.(2001)]{crowl01}
Crowl, H.H., Sarajedini, A., Piatti, A.E., Geisler, D., Bica, E., Clari\`a,
J.J., \& Santos, J.F.C, Jr. 2001, AJ, 122, 220

\bibitem[de Grijs et al.(2002)]{degrijs02}
de Grijs, R., Gilmore, G.F., Johnson, R.A., Mackey, A.D. 2002, MNRAS, 33, 245

\bibitem[de Grijs \& Anders(2006)]{degrijis06}
de Grijs, R., \& Anders, P. 2006, MNRAS, 366, 295

\bibitem[de Grijs \& Goodwin(2008)]{degrijs08}
de Grijs, R., \& Goodwin, S.P. 2008, MNRAS, 383, 1000

%\bibitem[Duchene, et al. (2001)]{duchene01}
%Duchene, G. , Simon, T. , Eisloffel, J., Bouvier, J. 2001, A\&A, 379, 147 

\bibitem[Eggleston(2006)]{eggleston06}
Eggleton P. 2006, {\it ``Evolutionary Processes in Binary and Multiple Stars''}. Cambridge Astrophys. No. 40. Cambridge, UK: Cambridge Univ. Press

\bibitem[Elmegreen(2007)]{elmegreen07}
Elmegreen, B.G. 2007, ApJ, 668, 1064

\bibitem[Elson et al.(1987)]{elson87}
Elson, R.A.W., Fall, S.M., \& Freeman, K. 1987, \apj, 323, 54

\bibitem[Fagotto et al.(1994)]{fagotto94}
Fagotto, F., Bressan, A., Bertelli, G., Chiosi, C. 1994, A\&AS, 104, 365

\bibitem[Fall \& Rees(1977)]{fall77}
Fall, S.M., \& Rees, M.J. 1977, 1977, MNRAS, 181, 37

\bibitem[Fall \& Zhang(2001)]{fall01}
Fall, S.M., \& Zhang, Q. 2001, \apj, 561, 751

\bibitem[Fall et al.(2009)]{fall}
Fall, S.M., Chandar, R., Whitmore, B.C. 2009, \apj, 704,453

\bibitem[Gieles \& Bastian(2008)]{gieles08}
Gieles, M., \& Bastian, N. 2008, A\&A, 482, 165

\bibitem[Gieles et al.(2010)]{gieles10}
Gieles, K., Sana, H., Portegiese Zwart, S.F. 2010, MNRAS, 402, 1750

\bibitem[Glatt et al.(2008b)]{glatt08b}
Glatt, K., Gallagher, J. S., III,glatt Grebel, E.K., Nota, A., Sabbi, E., Sirianni, M., Clementini, G., Tosi, M., Harbeck, D., Koch, A., Cracraft, M. 2008, \aj, 135, 1106

\bibitem[Glatt et al.(2008a)]{glatt08a}
Glatt, K., et al. 2008, \aj, 136, 1703

\bibitem[Gnedin \& Ostriker(1997)]{gnedin97}
Gnedin, O.Y., \& Ostriker, J.P. 1997, \apj, 474, 223

\bibitem[Goodwin(1987)]{goodwin97}
Goodwin, S.P. 1987, MNRAS, 286, 669

\bibitem[Goodwin \& Bastian(2006)]{goodwin06}
Goodwin, S.P., \& Bastian, N. 2006, MNRAS, 373, 752

\bibitem[Gouliermis et al.(2004)]{gouliermis04}
Gouliermis, D., Keller, S.C., Kontizas, M., Kontizas, E., Bellas-Velidis, I. 2004, A\&A, 416, 137

\bibitem[Hatzidimitriou et al.(1993)]{hatzidimitriou93}
Hatzidimitriou, D., Cannon, R. D., \& Hawkins, M. R. S. 1993,
MNRAS, 261, 873

%\bibitem[Hatzidimitriou et al.(2005)]{hatzidimitriou05}
%Hatzidimitriou, D., Stanimirovic, S., Maragoudaki, F., Staveley-Smith, L., Dapergolas, A., Bratsolis, E.2005, MNRAS, 360, 11

\bibitem[Hilditch et al.(2005)]{hilditch05}
Hilditch, R.W., Howarth, I.D., Harries, T.J. 2005, MNRAS, 357, 304

\bibitem[Hillenbrand(1997)]{hillenbrand97}
Hillenbrand, L.A. 1997, AJ, 114, 198

\bibitem[Hillenbrand \& Hartmann(1998)]{hillenbrand98}
Hillenbrand, L.A. \& Hartmann, L.W. 1998, ApJ, 492, 540

\bibitem[Hills(1980)]{hills80}
Hills, J.G. 1980, \apj, 235,986

\bibitem[Keller et al.(2000)]{keller00}
Keller, S.C., Bessell, M.S., Da Costa, G.S. 2000, AJ, 119, 1748

\bibitem[King(1966)]{king66}
King, I.R. 1966, \aj, 71, 276

\bibitem[Koekemoer et al.(2002)]{koekemoer02}
Koekemoer, A.M., Fruchter, A.S., Hook, R., Hack, W. 2002, in {\it Hubble} after the Installation of the ACS and the NICMOS Cooling System, ed. S. Arribas, A. Koekemoer, \& B. Whitmore (Baltimore: STScI), 337

\bibitem[Krist(2003)]{krist03}
Krist, J. 2003, ACS WFC \& HRC Field-dependent PSF variations Due to Optical and Charge Diffusion Effects (ACS/ISR/ 2003-06; Baltimore, MD:STScI)

\bibitem[Kupper et al.(2010)]{kuepper10}
Kupper, A.H., Kruppa, P., Baumgardt, H. Heggie, D.C. 2010, MNRAS, 407, 2241

\bibitem[Lada \& Lada(2003)]{lada03}
Lada, C.J, Lada, E.A. 2003, ARA\&A, 41, 57

\bibitem[Lah, et al.(2005)]{lah05}
Lah, P., Kiss, L. L., \& Bedding, T. R. 2005, MNRAS, 359, L42

\bibitem[Mathewson et al.(1988)]{mathewson88}
Mathewson, D. S., Ford, V. L., \& Visvanathan, N. 1988, ApJ, 333,
617

\bibitem[Matzner \& McKee(2000)]{matzner00}
Matzner, C.D., McKee, C.F. 2000, \apj, 545, 364

\bibitem[McCumber \& Garnett(2005)]{mccumber05}
McCumber, M.P., Garnett, D.R. 2005, \aj, 514, 96

\bibitem[Meylan \& Heggie(1997)]{meylan97}
Meylan, G. \& Heggie, D.C. 1997, A\&A Rev., 8, 1

\bibitem[Mizuno et al.(2001)]{mizuno01}
Mizuno, N., Rubio, M., Mizuno, A., Yamaguchi, R., Onishi, T., Fukui, Y. 2001, Publ. Astron. Soc. Jpn. 53, L45

\bibitem[Montegriffo et al.(1995)]{montegriffo95}
Montegriffo, P. Ferraro, F.R., Fusi Pecci, F., Origlia, L. 1995, MNRAS, 276, 739

\bibitem[No$\ddot{e}$l et al.(2007)]{noel07}
No$\ddot{e}$l, N.E.D., Gallart, C., Costa, E., \& M\'{e}ndez, R.A. 2007, \aj, 133, 2037

\bibitem[No$\ddot{e}$l et al.(2010)]{noel10}
No$\ddot{e}$l, N.E.D., Aparicio, A., Gallart, C., Hidalgo, S.L., Costa, E., \& M\'{e}ndez, R.A. 2010, \apj, 705, 1260

\bibitem[Nota et al.(2006)]{nota06} 
Nota, A., Sirianni, M., Sabbi, E., Tosi, M., Clampin, M., Gallagher, J.S., Meixner, M., Oey, M.S., Pasquali, A., Smith, L.J., Walterbos, R., Mack, J 2006, \apj, 640, L29 

\bibitem[Parisi et al.(2010)]{parisi10}
Parisi, M.C., Geisler, D., Grocholski, A.J., Sarajedini, A. 2010, \aj, 139, 1168

\bibitem[Piatti et al.(2007)]{piatti07}
Piatti, A.E., Sarajedini, A., Geisler, D., Clark, D., Seguel, J. 2007, MNRAS, 377, 300

\bibitem[Price \& Bate(2009)]{price09}
Price, D.J., Bate, M.R. 2009, MNRAS, 398, 33

%\bibitem[Priebisch et al.(1999)]{priebisch99}
%Priebisch, T. , Balega, Y. , Hofmann, K. , Weigelt, G., Zinnecker, H. 1999, New Astron., 4, 531

\bibitem[Sabbi et al.(2007)]{sabbi07}
Sabbi, E., Sirianni, M., Nota, A., Tosi, M., Gallagher, J., Meixner, M., Oey, M. S., Walterbos, R., Pasquali, A., Smith, L. J., Angeretti, L. 2007, AJ, 133, 44

\bibitem[Sabbi et al.(2008)]{sabbi08}
Sabbi, E., Sirianni, M, Nota, A., Tosi, M., Gallagher, J.S., Smith L.J., Angeretti, L., Meixner., M., Oey, M.S., Walterbos, R., Pasquali, A. 2008, \aj, 135, 173

\bibitem[Sabbi et al.(2009)]{sabbi09}
Sabbi, E., et al. 2009, \apj, 703, 721

\bibitem[Sagar \& Richtler(1991)]{sagar91}
Sagar, R., \& Richtler, T. 1991, A\&A, 250, 324

\bibitem[Salpeter(1955)]{salpeter}
Salpeter, E.E. 1955, \apj, 121, 161

\bibitem[Scalo(1986)]{scalo86}
Scalo, J. M. 1986, Fund. Cosmic Phys., 11, 1

\bibitem[Simon et al.(2007)]{simon07}
Simon, J.D., Bolatto, A.D., Whitney, B.A., Robitaille, T.P., Shah, R.Y., Makovoz, D., Stanimirovi\`{c}, S., Barb\`{a}, R.H., Rubio, M. 2007, \apj, 669, 327

\bibitem[Sirianni et al.(2002)]{sirianni02}
Sirianni, M., Nota, A., De Marchi, G., Leitherer, C., Clampin, M. 2002, \apj, 533, 203

\bibitem[Sirianni et al.(2005)]{sirianni05}
Sirianni, M., et al. 2005, PASP, 117,1049

\bibitem[Smith et al.(2011)]{Smith11}
Smith, R., Fellhauer, M., Goodwin, S., \& Assman, P. 2011, MNRAS, submitted (ArXiv:1102.5360)

\bibitem[Spitzer(1969)]{spitzer69}
Spitzer, L. 1969, ApJ, 127, 17

\bibitem[Spitzer(1987)]{spitzer87}
Spitzer, L. 1987, Dynamical Evolution of Globular Clusters, Princeton Univ. Press., Princeton, NJ

\bibitem[Sosynski et al.(2002)]{soszynski02}
Sosynski, I., et al. 2002, Acta Astron., 52, 369

\bibitem[Stolte et al.(2002)]{stolte02}
Stolte, A., Grebel, E.K., Brandner, W., Figer, D.F. 2002, A\&A, 394, 459

\bibitem[Tutukov(1978)]{tutukov78}
Tutukov, A.V. 1978, A\&A, 70, 57

\bibitem[Vesperini(1997)]{vesperini97}
Vesperini, E. 1997, MNRAS, 287, 915

\bibitem[Vesperini(1998)]{vesperini98}
Vesperini, E. 1998, MNRAS, 299, 1019

\bibitem[Vesperini \& Zepf(2003)]{vesperini03}
Vesperini, E., \& Zepf, S.E. 2003, \apj, 587, L97

\bibitem[Weidner, Kroupa \& Maschberger(2009)]{weidner09}
Weidner, C., Kroupa, P., \& Maschberger, T. 2009, MNRAS, 393, 663

\bibitem[Whitmore et al.(2007)]{withmore05}
Whitmore, B.C., Chandar, R., Fall, S.M. 2007, \aj, 133, 1067

\bibitem[Whitworth(1979)]{whitworth79}
Whitworth, A. 1979, MNRAS, 186, 59

%\bibitem[Zaritsky et al.(2000)]{zaristky00}
%Zaritsky, D. Harris, J., Grebel, E.K., Thompson, I.B.2000, \apj, L53

\bibitem[Zaritky et al.(2002)]{zaritsky02}
Zaritsky, D. Harris, J., Thompson, I.B, Grebel, E.K, Massey, P. 2002, \aj, 123, 855

%\bibitem[Zaritsky \& Harris(2004)]{zaritsky04}
%Zaritsky, D., \& Harris, J. 2004, \apj, 604, 167 

\bibitem[Zhang \& Fall(1999)]{zhan99}
Zhang, Q. \& Fall, S.M. 1999, \apj, 527, L81
\end{thebibliography}
\end{document}